\documentclass[prd,showpacs,preprint,preprintnumbers,amsmath,amssymb,eqsecnum,nofootinbib]{revtex4-1}
\usepackage{graphicx,amssymb,amsmath,amsthm,amsfonts,mathrsfs,epsfig,epsf}
\usepackage{dcolumn}
\usepackage{amsfonts,mathrsfs}
\usepackage{bm}
\usepackage{subfigure}
\usepackage{color}
\usepackage{url}
\usepackage{hyperref}
\hypersetup{
setpagesize=false,
 bookmarksnumbered=true,%
 bookmarksopen=true,%
 colorlinks=true,%
 linkcolor=blue,
 citecolor=red,
}

\usepackage[top=20truemm,bottom=20truemm,left=20truemm,right=20truemm]{geometry}

\newcommand{\D}{{\rm d}}
\newcommand{\eq}[1]{\mbox{Eq.~(\ref{#1})}}
\newcommand{\fig}[1]{\mbox{Fig.~\ref{#1}}}

\begin{document}


\title{Confined Penrose process with charged particles}

\author{Takafumi Kokubu} 
\email{kokubu@hunnu.edu.cn, 14ra002a@al.rikkyo.ac.jp}
\author{Shou-Long Li}
\email{shoulongli@hunnu.edu.cn}
\author{Puxun Wu} 
\email{pxwu@hunnu.edu.cn}
\author{Hongwei Yu} 
\email{hwyu@hunnu.edu.cn}
\affiliation{Department of Physics and Synergetic Innovation Center for Quantum Effects and Applications, Hunan Normal University, Changsha, Hunan 410081, China}

\date{\today}

\begin{abstract}
We show that kinematics of charged particles allows us to model the growth of particles' energy by consecutive particle splits, once a spherical mirror as a perfectly reflective boundary is placed outside a charged black hole.
We consider a charged version of the Penrose process, in which a charged particle decays into two fragments, one of them has negative energy and the other has positive energy that is larger than that of the parent particle.
The  confinement system with the mirror makes the particles' energy amplified each time a split of the parent particle occurs. Thus, the energy is a monotonically increasing function of time. However, the energy does not increase unboundedly, but rather asymptotes to a certain finite value, implying no instability of the system in this respect.

\end{abstract}


\maketitle


\section{Introduction}
Black holes (BHs) definitely play a leading role in gravitational physics as a representative of strong gravitational fields. The physics of BHs 
has revealed a number of 
surprising features about the nature of spacetime, 
 for instance, the possibility of extracting energy from BHs, which were believed to swallow anything around. This fundamental and important issue has been the focus of attention for many years, as energy extraction or amplification mechanisms using BHs are expected to explain the origins of highly energetic particles and relativistic jets \cite{Blandford:1977ds, Banados:2009pr, Banados:2010kn, Blandford:2017chu, Brito:2015oca, Schnittman2018}.

Penrose first 
demonstrated the possibility of energy extraction from BHs with an explicit example.  If the test particle splits into two fragments within the Kerr BH ergo region, then one fragment can have a negative energy (measured by an observer at infinity) \cite{Penrose1969, PenroseFloyd1971}  and  the other fragment has more energy than the  parent particle  
as a result of energy conservation. This mechanism of energy extraction is called the ``Penrose process".

``Superradiance" in gravitational physics is another example of energy extraction, which was first proposed by Zel'dovich \cite{Zeldovich1971}. The amplitude of waves reflected by a rotating cylinder is greater than the amplitude of the incident wave under certain conditions.
It was shown in Ref.~\cite{PressTeukolsky1972} that superradiance also occurs in the Kerr BH spacetime. The scattering of the massless test scalar field incident on the Kerr BH with a frequency lower than a threshold was calculated. It was confirmed that scattered amplitudes are larger than the incident ones.

In the above energy amplifying mechanisms, the existence of the angular momentum of BHs is necessary.
However, it is also known that even spherically symmetric BHs provide a chance to amplify energies  of test particles and fields. 
The charged test particles around a spherically symmetric charged BH, e.g., the Reissner--Nordstr\"om BHs, have Coulomb energy in addition to gravitational energy. It is known that when the sign of particle's charge is different from that of the Reissner--Nordstr\"om BHs, there can be a region where the particle's energy becomes negative near the BHs \cite{DenardoRuffini1973}. The origin of this region is clearly not from angular momentum but from charge. This region is called the generalized ergo region (or effective ergo region). 
If the split of a positively charged particle occurs in this region and one of fragments has a negative charge that will be swallowed into the positively charged BH, 
then the other fragment escapes to infinity at greater energy than the incident one. 
Such  ``a charged Penrose process'' occurs even in a flat spacetime. See Refs. \cite{DenardoTreves1973, Dadich1980, Zaslavskii:2018ngs}.
It should be emphasized that the energy 
amplification by the original Penrose process in the Kerr spacetime is severely limited \cite{Chandrasekhar1985} (
($E_{in}-E_{out})/E_{in}  \sim$  21 \%, with energies of an incident particle $E_{in}$ and a decayed fragment $E_{out}$), while the process in charged BHs has no upper limit. The larger the negative charge swallowed by the charged BHs, the more energetic the escaping particles.
There is also an electrical counterpart for superradiance of scalar fields. The setup is similar to that of the Kerr, but a test charged scalar field placed around the Reissner--Nordstr\"om BH allows the amplitudes of reflected waves to be larger than that of the incident ones, when the frequency of the field is lower than a threshold value.

An interesting application of superradiance was  also made in the study by Press and Teukolsky in Ref. \cite{PressTeukolsky1972}, which they named the ``BH bomb".
The BH bomb is a continuous energy-amplifying phenomenon with the aid of a perfectly reflective boundary placed outside the horizon, so that the reflected wave with a larger amplitude increased by superradiance is further reflected by the reflective boundary. This reflected incident wave will be scattered again, causing even larger amplitudes.  
As a result, the amplitude of the scalar field increases exponentially with time, indicating an instability of the system.
If massive scalar fields are on the Kerr BHs, the mass-term in the wave equation naturally provides a reflecting potential barrior, causing a BH bomb without an artificial boundary \cite{Damour:1976kh}. A negative cosmological constant also provides a natural confinement~\cite{Hawking:1999dp}.

In addition, a charged counterpart of BH bombs was also discovered. The system of charged massless scalar fields with the  superradiance condition around the Reissner--Nordstr\"om BHs surrounded by a reflective mirror shows an exponential growth of the amplitude of the field \cite{Bekenstein:1973mi}.

As explained so far, energy amplification mechanisms have been found in both test particles and test fields.
However, to the best of our knowledge, a continuous energy amplifying mechanism in the particle picture, like the BH bombs by scalar fields, has not been explored in detail~\footnote{There is a study on cascade of Penrose process around ``stars'' with ergo region, indicating an ergo region instability, see Ref. \cite{Vicente:2018mxl}.}.
Therefore, in this paper we ask whether BH bombs, or a similar continuous energy amplification phenomenon, occur when test particles are confined around BHs, and if so, does it show an exponential energy growth that indicates instability?
To answer these questions, this paper focuses on charged particles that move around the Reissner--Nordstr\"om BH.
In Ref.~\cite{DenardoRuffini1973}, the explicit charged Penrose process by charged particles with angular momentum is shown with fixed parameters.  In the present paper, in contrast, the charged Penrose process by charged particles without angular momentum, i.e., charged particles in radial motion,  will be considered with general parameters. This is in marked contrast to the previous study by Ref.~\cite{DenardoRuffini1973}.

This paper is organized as follows. In Sec. \ref{sec:EOM}, we prepare equations of motion of charged test particles in the (sub-)extremal Reissner--Nordstr\"om spacetime. 
In Sec. \ref{sec:PP}, we introduce the charged Penrose process by radially moving charged particles and propose the Penrose process in a confined system by putting a reflective boundary outside BHs.
 Sec. \ref{sec:summary} is devoted to  the summary and conclusion.
We take units of $c=G=1$ throughout the paper.

\section{Equation of motion}
\label{sec:EOM}
In this section, we derive the equation of motion for a charged test particle.
With the potential 1-form, $A_\mu \D x^\mu=-(\mathcal Q/r)\D t$, a static and spherically symmetric black hole with charge $\mathcal Q$ uniquely leads to the Reissner--Nordstr\"om solution of the Einstein's equations. The  spacetime metric of the solution is written by
\begin{align}
\D s^2=-f(r)\D t^2+f(r)^{-1}\D r^2+r^2( \D \theta^2+\sin^2\theta\D \phi^2)
\label{ds}
\end{align} 
with
\begin{align}
 f(r)=&1-2M/r+\mathcal Q^2/r^2.
\end{align}
The parameter $M$ is the gravitational mass and $\mathcal Q$ is the charge of the black hole. 
We assume that $M>0$ and $\mathcal Q> 0$ without loss of generality.
The radius of the event horizon, $r_H$, is the larger root of the equation $f(r_H)=0$, which is $r_H=M+\sqrt{M^2-\mathcal Q^2}$.

The Lagrangian for a charged test particle is written as
\begin{align}
\mathcal L=\frac{g_{\alpha\beta}}{2}\dot x^\alpha \dot x^\beta+\tilde qA_\alpha \dot x^\alpha,
\end{align}
where $\tilde q$ is the particle's charge, $\lambda$ the affine parameter, and $\dot X:=\D X/\D \lambda$. Here $\lambda$ is related to the particle's proper time as $\tau=m \lambda$ with the particle's rest mass $m$.
From the four-momemtum of the particle, $p^\mu:=\dot x^\mu$,
\begin{align}
p^t=\dot t=\frac{1}{f}\left(\tilde E-\frac{\tilde q \mathcal Q}{r}\right), ~~p^r=\dot r;,
\end{align}
where $\tilde E$ is the particle's energy that is measured at infinity. 
Since $\dot t$ should be positive, the inequality,
\begin{align}
\tilde E-\tilde q \mathcal Q/r>0, \label{forward-in-time}
\end{align}
must hold outside the event horizon.
Here, we review energetics of the particle. By solving inequality (\ref{forward-in-time}) in terms of $r$,  we obtain, depending upon the sign of energy $\tilde E$,
\begin{align}
r>r_E:=\tilde q \mathcal Q/\tilde E \quad &{\rm for} \quad \tilde E>0, \nonumber  \\
r<r_E \quad &{\rm for} \quad \tilde E<0. \label{var-forward-in-time}
\end{align}
In the latter case in inequality (\ref{var-forward-in-time}), particles with $\tilde E<0$ can exist within the region $r>0$
 only if $\tilde q \mathcal Q<0$.
Thus, negative energy particles can exist in the region
\begin{align}
r_H<r<r_E, \label{Gergo}
\end{align}
if 
\begin{align}
r_H<r_E=|\tilde q \mathcal Q|/|\tilde E|. \label{amp-condition1}
\end{align}
The region of \eq{Gergo} is the generalized ergo region.
Inequality (\ref{amp-condition1}) is an analog to the super-radiance condition for charged scalar fields, $r_H<q\mathcal Q/\omega$, where $\omega$ is frequency, and $q$ the charge of the field \cite{Gibbons1975}.
The generalized ergo region can extend outside the horizon if $r_E>r_H$.
It is noted that the generalized ergo region is not exclusively a characteristic of the BH, as is the case in the more common Kerr's ergo region where this region purely intrinsic to the background spacetime. In the present context, the generalized ergo region depends also on the mass and charge of the {\it particle}.

In this paper, we restrict our attention to the radial motion of particles for two reasons. One is for simplicity, and the other is that the range of the generalized ergo region decreases as particle's angular momentum increases, implying the allowed region of negative energy state is the widest for radial particles. 
As a result, energy extraction by radial motions are easier than that by non-radial ones.
See Ref.~\cite{DenardoRuffini1973} for this point.
Then, we have the energy equation for the radially moving particle from the normalization of the four-momemtum, $-m^2=p^\alpha p_\alpha$, as
\begin{align}
\dot r^2+V(r)=0
\end{align}
with
\begin{align}
V(r)=m^2 f(r)-\left(\tilde E-\frac{\tilde q \mathcal Q}{r}\right)^2. \label{potential}
\end{align}
In this paper we consider the parameter region of $M\geq \mathcal Q>0$ that describes sub-extremal ($M>\mathcal Q$) or extremal ($M=\mathcal Q$) BHs.
The effective potential \eq{potential} has, in general, two zero points $r_{0\pm}$, satisfying $V(r_{0\pm})=0$, which is solved as
\begin{align}
r_{0\pm}=\frac{M}{E^2-1}\left(Eq Q-1\pm \sqrt{(Eq Q-1)^2-(E^2-1)(q^2-1) Q^2} \right). \label{r0}
\end{align}
Here we have defined 
\begin{align}
  Q:=\mathcal Q/M, \quad E:=\tilde E/m, \quad q:=\tilde q/m.
\end{align}
Note that $r_{0+}>r_{0-}$ if $E^2>1$ and $q^2>1$.

From the fact that the inside of the root in  \eq{r0} should be non-negative if turning points exist, the charge $q$ has a lower bound as
\begin{align}
q\geq  \frac{E+\sqrt{(E^2-1)(1-Q^2)}}{Q}. \label{qc}
\end{align}
The equality of \eq{qc} holds when $(Eq Q-1)^2-(E^2-1)(q^2-1) Q^2=0$, satisfying the static condition, $V(r)=\D V(r)/\D r=0$.

It can be shown by straightforward calculations that particles coming from infinity can bounce back at the outer turning point, $r_{0+}$, laying outside the horizon, if and only if
\begin{align}
q> \frac{E+\sqrt{(E^2-1)(1-Q^2)}}{Q} \quad {\rm and} \quad E>1. \label{bounce-condition}
\end{align}
For other choices of $q$ and $E$, incoming particles just fall into the BH horizon. 

In the original Penrose process, a parent particle is supposed to decay (split) at a turning point. Here, we assume that Particle $0$ satisfying \eq{bounce-condition} is the one to decay.
Thus, the turning point of radius $r_s$, where the splitting of the parent particle occurs, is $r_s=r_{0+}$.

\section{Confined Penrose process}\label{sec:PP}
In this section we first introduce the charged Penrose process by charged particles that move along the radial direction. The Penrose process using charged particles was first implemented by Denardo and Ruffini in Ref.~\cite{DenardoRuffini1973}, where a single explicit example was given by particles in spiral motion with a given  set of numerical values for the rest mass, energy,  charge, and  angular momentum of the particles, as well as the charge of the background. Here, we will consider the energy extraction process by radially moving particles with general  parameters that characterize the process.
We will also show that the charged Penrose process occurs many times by placing a reflective wall outside the horizon.
As a consequence, the energy of the particle increases with time by  the repeating process in the confined system.
 
Imagine that a charged particle, Particle 0, decays into two new charged particles, Particle 1 and 2, at a split (decay) radius $r_s$.
The potential for Particle $A (=0,1,2)$ is given by
\begin{align}
\dot r_A^2+V_A(r_A)=0 \quad {\rm with} \quad V_A(r_A)=m_A^2 f(r_A)-\left(\tilde E_A-\frac{\tilde q_A \mathcal Q}{r_A}\right)^2. 
\label{potential-A}
\end{align}
Quantities with subscript $A$ belong to Particle $A$.

\begin{figure}[htbp]
  \begin{center}
          \includegraphics[scale=0.7]{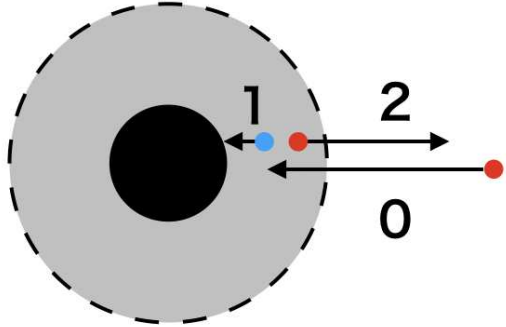}
    \caption{Schematic picture for `charged' Penrose process. Positively charged particle, Particle $0$, splits into two oppositely charged particles, Particle $1$ and $2$, within a generalized ergo region (gray region) of the BH at the center (black region).
\label{fig-PP}}
  \end{center}
\end{figure}

If Particle $0$ is to have an outer turning point, from the condition \eq{bounce-condition},
\begin{align}
q_0>q_c:= \frac{E_0+\sqrt{(E_0^2-1)(1-Q^2)}}{Q}
 \quad {\rm and} \quad E_0>1, \label{0-bounce-condition}
\end{align}
must hold.
We assume that Particle 1 has a negative charge, while Particle 2 has a positive charge. See \fig{fig-PP} for a schematic picture of the process.
In this case, at the split radius, the momentum conservation is written as $p^\mu_0=p^\mu_1+p^\mu_2$.
The $t$-component of the conservation is
\begin{align}
\tilde E_0-\frac{\tilde q_0 \mathcal Q}{r}=\tilde E_1-\frac{\tilde q_1 \mathcal Q}{r}+\tilde E_2-\frac{\tilde q_2 \mathcal Q}{r}, 
\label{t-momentum}
\end{align}
whereas the $r$-component is 
\begin{align}
0=\dot r_1+ \dot r_2. \label{r-momentum}
\end{align}
Meanwhile, there is a charge conservation:
\begin{align}
\tilde q_0=\tilde q_1+\tilde q_2. \label{q-cons}
\end{align}
From \eq{t-momentum} and \eq{q-cons}, we obtain the energy conservation,
\begin{align}
\tilde E_0=\tilde E_1+\tilde E_2. \label{energy-cons}
\end{align}
We denote the rest mass of Particle $i (=1,2)$ after the split as $m_i$, which is
\begin{align}
m_i=\alpha_i m_0,
\label{mass}
\end{align}
where the proportional constant is given by $0<\alpha_i<1$ and $\alpha_1+\alpha_2<1$.
From \eq{mass} and \eq{energy-cons}, we have
\begin{align}
E_0=\alpha_1 E_1+\alpha_2E_2. \label{energy-cons-re}
\end{align}
Similarly, the charge conservation reduces to
\begin{align}
q_0=\alpha_1q_1+\alpha_2q_2. \label{q-cons-re}
\end{align}
We  define the energy amplification factor $\eta$ as
\begin{align}
\eta:=\frac{{\rm positive~ energy~ of~ decayed ~  fragment}}{{\rm incident ~ energy}}=\frac{\tilde E_2}{\tilde E_0}=\frac{\alpha_2 E_2}{E_0}=1-\frac{\alpha_1 E_1}{E_0}. \label{def-eta}
\end{align}
Equation (\ref{def-eta}) shows $\eta>1$ if and only if $E_1<0$.

For simplicity, we assume that two fragments are at rest at first \footnote{We will discuss  this assumption in the last section.}, i.e.,
\begin{align}
\dot r_1(r_s)=\dot r_2(r_s)=0. \label{rest}
\end{align}
Then \eq{r-momentum} is trivially satisfied.
Now, with the aid of $\dot r_{i}(r_s)=0$ and \eq{potential-A}, the energy $E_i$ for Particle $i$ is uniquely determined to be
\begin{align}
E_i=\frac{q_iQ}{x_s}+\sqrt{f(x_s)}. \label{Ei}
\end{align}
where we defined $x:=r/M$ and $x_s:=r_s/M$.
Although $\dot r_{i}(r_s)=0$ has in general two solutions in terms of $E_i$, we excluded the negative root solution because it does not satisfy $\D t/\D \lambda>0$ outside the horizon.

\begin{figure}[htbp]
  \begin{center}
    \begin{tabular}{c}

      \begin{minipage}{0.5\hsize}
        \begin{center}
          \includegraphics[scale=0.65]{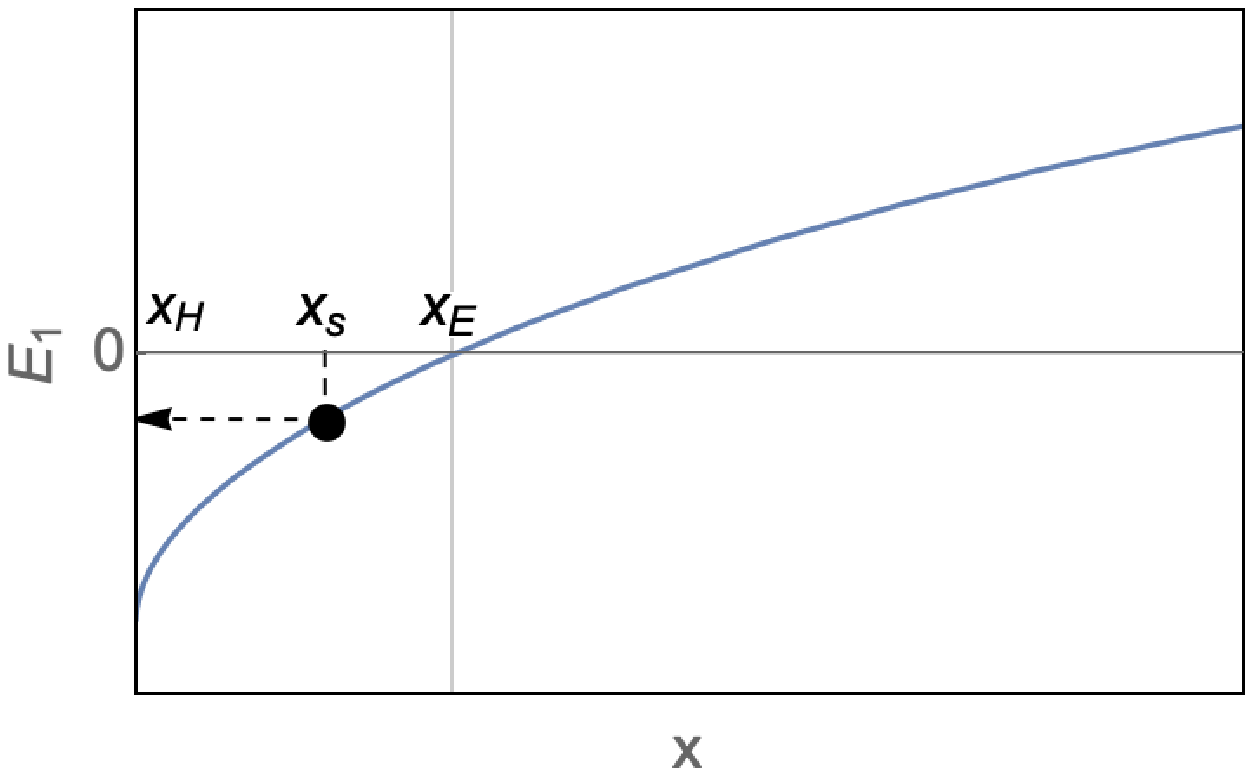}
          (a) 
        \end{center}
      \end{minipage}
      
      \begin{minipage}{0.5\hsize}
        \begin{center}
          \includegraphics[scale=0.65]{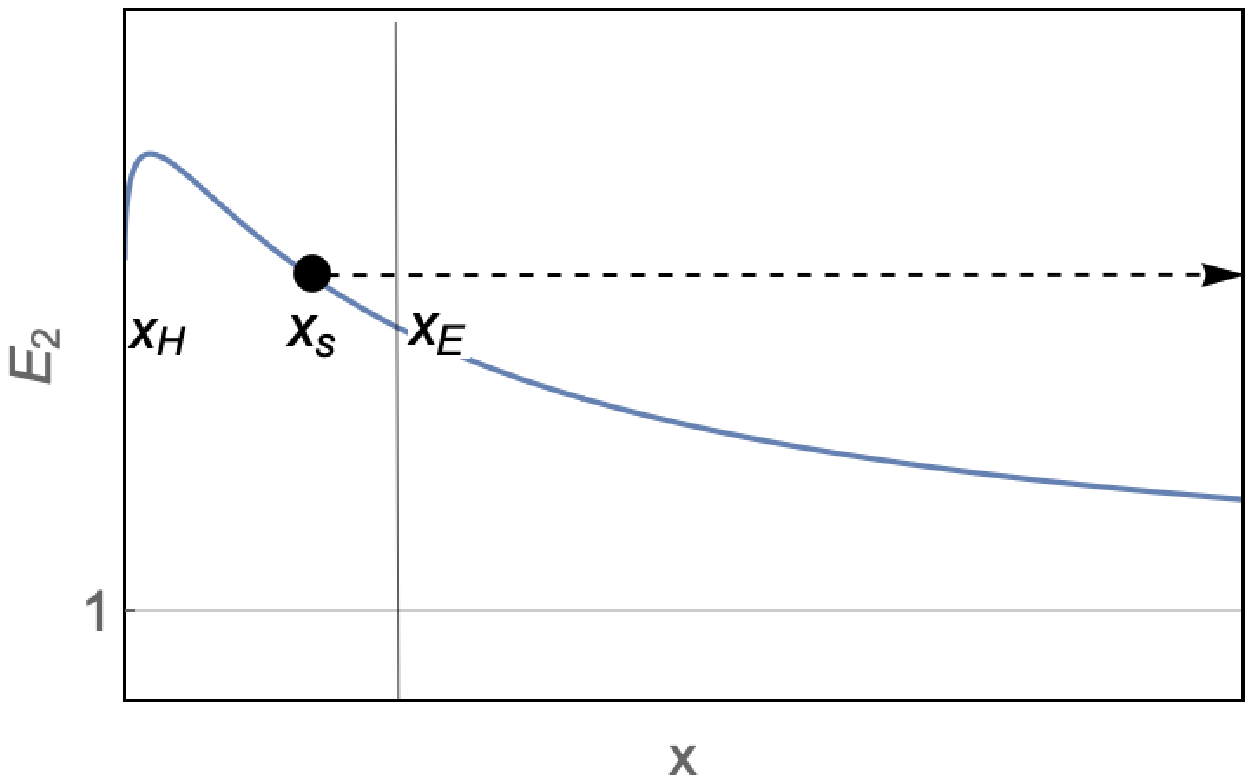}
         (b) 
        \end{center}
      \end{minipage}
 
\\
    \end{tabular}
    \caption{
    Schematic figure of $g_i$. The horizontal dotted lines denote values of the energy for Particle $i$.
 (a)  $g_1(x)$ for Particle 1 with negative charge. Fragment's energy must be negative if the split occurs inside the generalized ergo region (shaded range), $x_H<x_s<x_E$. Since Particle 1 moves in the range $x\leq x_s$, it cannot escape to infinity, but falls into the BH. (b) $g_2$ for Particle 2 with positive charge. Particle's energy must be positive and escapes outwardly.}    
\label{fig-P12}
  \end{center}
\end{figure}

We note that a variant expression of  the allowed range for $x$ of particle motions, $V_i(x) \leq 0$, can be written as
\begin{align}
E_i\geq \frac{q_iQ}{x}+\sqrt{f(x)}=:g_i(x). \label{g}
\end{align}
The generalized ergo sphere, the boundary of the generalized ergo region is given by $g_1(x_E)=0$, which is written as $x_E=1+\sqrt{1+(q_1^2-1)Q^2}$.
Here, we mention some general properties for Particle $i$. 
For Particle 1 with negative energy, we have $\D g_1/\D x=-q_1Q/x^2+(2\sqrt{f})^{-1}\D f/\D x$ from \eq{g}.
Since $\D f/\D x>0$ for $x>x_H$, $g_1$ is a monotonically increasing function of $x$ outside the horizon if $q_1Q<0$.
Thus, Particle 1  must move inwardly and  fall into the horizon.
See \fig{fig-P12} (a) for $g_1$. The generalized ergo region is given as a shaded region in the figure.
For Particle $2$, its energy is apparently positive from \eq{g}. See \fig{fig-P12}  for schematic figure of $g_i(x)$.

Now, we explain the rest mass is conserved before and after the decay event. From \eq{energy-cons} and \eq{Ei}, $x_s$ must satisfy
\begin{align}
E_0
=\frac{q_0Q}{x_s}+(\alpha_1 +\alpha_2 )\sqrt{f(x_s)}.
\label{energyATxs}
\end{align}
Meanwhile, Particle $0$ should satisfy $\dot r_0=0$, i.e.,
\begin{align}
E_0=\frac{q_0Q}{x_s}+\sqrt{f(x_s)}, \label{energyATxs2}
\end{align}
at the split radius $x_s$. From \eq{energyATxs} and  \eq{energyATxs2}, we find that if the split is to happen,  the relation, $\alpha_1 +\alpha_2 =1$, must hold.
This relation clearly shows that no mass-defect of Particle $0$ is allowed during split, i.e., particle's rest mass is conserved,
\begin{align}
m_0=m_1+m_2. \label{m-cons}
\end{align}
\eq{m-cons} does not hold however for more general situations with non-vanishing velocity, $\dot r_i (r_s) \neq 0$ \cite{Dadich1980}.

To have $E_1<0$, besides \eq{amp-condition1}, it is sufficient that the split occurs 
within the generalized ergo region:
\begin{align}
x_s<x_{\rm E}.  \label{sufficient-condition}
\end{align}
Solving inequality (\ref{sufficient-condition}) in terms of the charge $q_1$ of Particle $1$, we obtain
\begin{align}
&x_s<x_{\rm E} \Leftrightarrow x_s<1+\sqrt{1+(q_1^2-1)Q^2} \nonumber \\
&\Rightarrow (x_s-1)^2<1+(q_1^2-1)Q^2 
\quad \Leftrightarrow \quad q_1^2Q^2>x_s^2-2x_s+Q^2 \nonumber \\
&\Rightarrow q_1<-\frac{x_s\sqrt{f(x_s)}}{Q}=:-q_{{\rm min}}. \label{qmin}
\end{align}
The negative sign before $q_{{\rm min}}$ in \eq{qmin} is necessary due to negativity of $q_1$.
Thus,  the necessary and sufficient condition for energy extraction under the present setup results from a combination of inequalities (\ref{amp-condition1}) and (\ref{sufficient-condition}), 
\begin{align}
x_H<x_s<x_E. \label{full-amp-condition}
\end{align}
As can be seen in the condition (\ref{qmin}), there is a lower bound for the absolute value of negative charge, $|q_1|$. The value of $|q_1|$, that is required for energy extraction, depends upon the value of $f$ in $q_{{\rm min}}$: $|q_1|$ becomes arbitrarily close to zero if the decay radius $x_s$ can be arbitrarily close to the horizon because $\lim_{x_s \to x_H}f(x_s)=0$. 
However, an arbitrarily small $|q_1|$ is in general unachievable, unless either Particle $0$ has a large amount of energy from the very beginning or the background spacetime is originally extremal. We will explain the reason behind by evaluating the minimum of $x_s$.
When we regard $x_s$ as a function of $q_0$, its derivative is given by
\begin{align}
\frac{\partial x_s}{\partial q_0}=\frac{Q^2}{E_0^2-1}\left(E_0+\frac{q_0Q-E_0}{\sqrt{(E_0q_0 Q-1)^2-(E_0^2-1)(q_0^2-1) Q^2} }\right). \label{dxsdq0}
\end{align}

\begin{figure}[htbp]
  \begin{center}
          \includegraphics[scale=0.7]{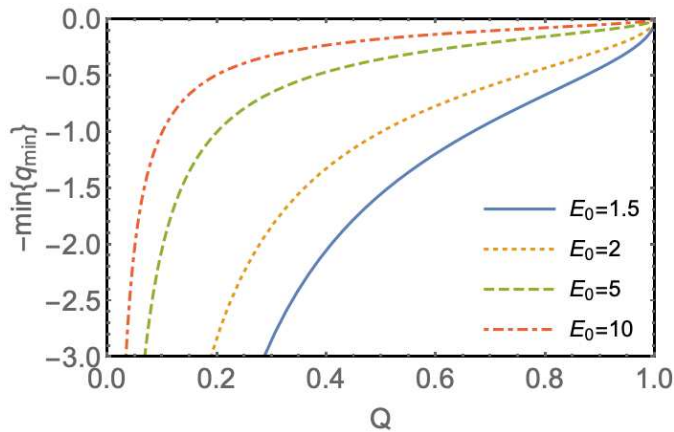}
    \caption{ Minimum of $q_{{\rm min}}$ as a function of the background charge $Q$. The Value of a negative charge $q_1$ that is supposed to be dropped into BHs must be below these lines in order to obtain negative $E_1$.
\label{fig-qminmin}}
  \end{center}
\end{figure}

On the other hand, from \eq{qc}, $q_0Q-E_0\geq q_cQ-E_0=\sqrt{(E_0^2-1)(1-Q^2)}\geq 0$. This proves  monotonicity of \eq{dxsdq0}, i.e., $\partial x_s/\partial q_0>0$ if $E_0>1$ and $q_0\geq q_c$. Hence, the minimum of $x_s$ is
\begin{align}
{\rm min}\{x_s\}=x_s|_{q_0=q_c}=1+\frac{E_0}{\sqrt{E_0^2-1}}\sqrt{1-Q^2}\geq 1+\sqrt{1-Q^2}=x_H.
\end{align}
This clearly shows that minimum of $x_s$ can be close to $x_H$ if either the background becomes the extremal ($\mathcal Q \to M$) or the particle is initially highly energetic $E_0 \to \infty$. 
As a consequence, $q_{{\rm min}}$ cannot be in general arbitrarily close to zero. Then, the minimum of $q_{{\rm min}}$ is
\begin{align}
{\rm min}\{q_{{\rm min}}\}=q_{{\rm min}}|_{q_0=q_c}=\frac{1}{Q}\sqrt{\frac{1-Q^2}{E_0^2-1}}. \label{qminmin}
\end{align}
This proves that the absolute value of the charge of the particle that falls into the BH  cannot be in general arbitrarily small.
We plot \eq{qminmin} in \fig{fig-qminmin}.
At this moment, we also found the necessary and sufficient condition \eq{full-amp-condition} in an explicit form as
\begin{align}
E_0>1, \quad  q_0>q_c, \quad {\rm and} \quad q_1<-q_{min}.
\end{align}

Let  us now evaluate the energy and charge that the fragments possess.
We take the initial charge for Particle $0$ as
\begin{align}
q_0=(1+\delta)q_c, \quad \delta >0.
\end{align}
We do not include the case with $\delta=0$, in which Particle $0$ results in an unstable  static solution at the split radius $x_s$, thus implying that Particle $2$ is also static so that it does not escape to infinity.
Also, we assume that 
the charge of particle $1$ is taken as the deviation from the lower limit as follows:
\begin{align}
q_1=-(1+\Delta)q_{{\rm min}}, \quad \Delta>0. \label{q1}
\end{align}
Then, the (negative) energy  for Particle 1 is, by substituting \eq{q1} into \eq{Ei}, given by
\begin{align}
E_1&=-\Delta \sqrt{f(x_s)}. \label{E1}
\end{align}
It is important to note that, for sub-extremal BHs, the energy of Particle 1 can be arbitrarily close to zero if we take the limit $\Delta \to 0$. 
On the other hand, the charge of the particle cannot be close to zero, even if the same limit is taken.
For extremal BHs, however, the energy as well as the charge can be arbitrarily close to zero even if we do not take the limit $\Delta \to 0$, because we can take the limit $x_s \to x_H$ only in the extremal case. 

Then, the energy and the charge for Particle 2 having positive energy, from Eqs. (\ref{energy-cons-re}), (\ref{q-cons-re}), (\ref{q1}) and (\ref{E1}), reduce to
\begin{align}
E_2&=\left(E_0+\alpha_1 \Delta \sqrt{f(x_s)}\right)/\alpha_2, \label{E2}\\
q_2&=\left\{ (1+\delta)q_c+\alpha_1(1+\Delta)q_{{\rm min}}\right\}/\alpha_2. \label{q2}
\end{align}
Note that $\tilde E_2=\alpha_2 E_2$ and $\tilde q_2=\alpha_2 q_2 $ become large when $\alpha_1 \to 1$, i.e., when the escaping particle has a very small mass.
It is immediate from Eqs. (\ref{E2}) and (\ref{q2}) that $E_2>E_0/\alpha_2>E_0>1$ and $q_2>(1+\delta)q_c/\alpha_2>q_c$. As a result, we have
\begin{align}
q_2>q_c \quad {\rm and} \quad E_2>1. \label{2-bounce-condition}
\end{align}
Comparing \eq{0-bounce-condition} and \eq{2-bounce-condition}, we notice that Particle 2 has qualitatively the same effective potential as Particle 0. We emphasize this feature is crucial for realizing consecutive Penrose process in a confined system. This point will be discussed below soon.

From Eqs. (\ref{def-eta}) and (\ref{E2}), the energy amplification factor is calculated as 
\begin{align}
\eta=1+\frac{\alpha_1 \Delta \sqrt{f(x_s)}}{E_0}. \label{eta}
\end{align}
When does $\eta$ take the maximum? We find that the choice of $E_0, \alpha_1$ and $\delta$ cannot increase $\eta$ drastically, because $E_0$ cannot be less than unity under the present setup, $\alpha_1$ is at most unity, and, $\delta$, which is included in $x_s$ via $q_0$, controls $f(x_s)$ which is at most unity.
Nonetheless, $\eta$ is unbounded, since $\Delta$ takes any value in the test particle situation.
It is recognized that, from the form of \eq{eta}, $\eta$ takes the maximum when $\alpha_1 \to 1$ with other parameters fixed. This means that the process is the most effective when  Particle $2$ is a photon. This feature is same as the Penrose process in the Kerr BHs \cite{Brito:2015oca}.

We have explained the charged Penrose process for radially moving particles.
Let us repeat the process implemented above but adding a reflective spherical wall outside the horizon.
The position of the wall is arbitrary. We place Particle $0$ between the horizon and the wall.
Particle $0$ satisfying $q_0>q_c$ and $E_0>1$ decays into Particle $1$ and $2$ at $x=x_s$. Particle 1 with negative energy falls to the BH,  and Particle 2 moving outwardly reflects off the wall and falls inward again.
As mentioned, since Particle 2 has {\it qualitatively the same effective potential} as Particle 0, it reaches the same $x_s$ again. 
Then, an another new decay can occur there as the first time. That is, Particle 2 decays into Particle 3 and 4 with  $m_3=\alpha_1m_2$ and $m_4=\alpha_2 m_2$ (we also assume that two fragments are initially at rest). We assume $-q_3=(1+\Delta)q_{{\rm min}}$ for Particle 3 with negative energy. This assumption means that the same amount of negative energy is produced at each decay event.
Under the present setup, $q_{{\rm min}}(x_s)$ is invariant because $x_s$ is invariant at each decay. Then, the energy of Particle 3 is $E_3=-\Delta \sqrt{f(x_s)}$, which is the same as Particle 1. Particle 4 satisfies $q_4>q_c$ and $E_4>1$ for the same reason for Particle 2. 

\begin{figure}[htbp]
  \begin{center}
    \begin{tabular}{c}

      \begin{minipage}{0.48\hsize}
        \begin{center}
          \includegraphics[scale=0.63]{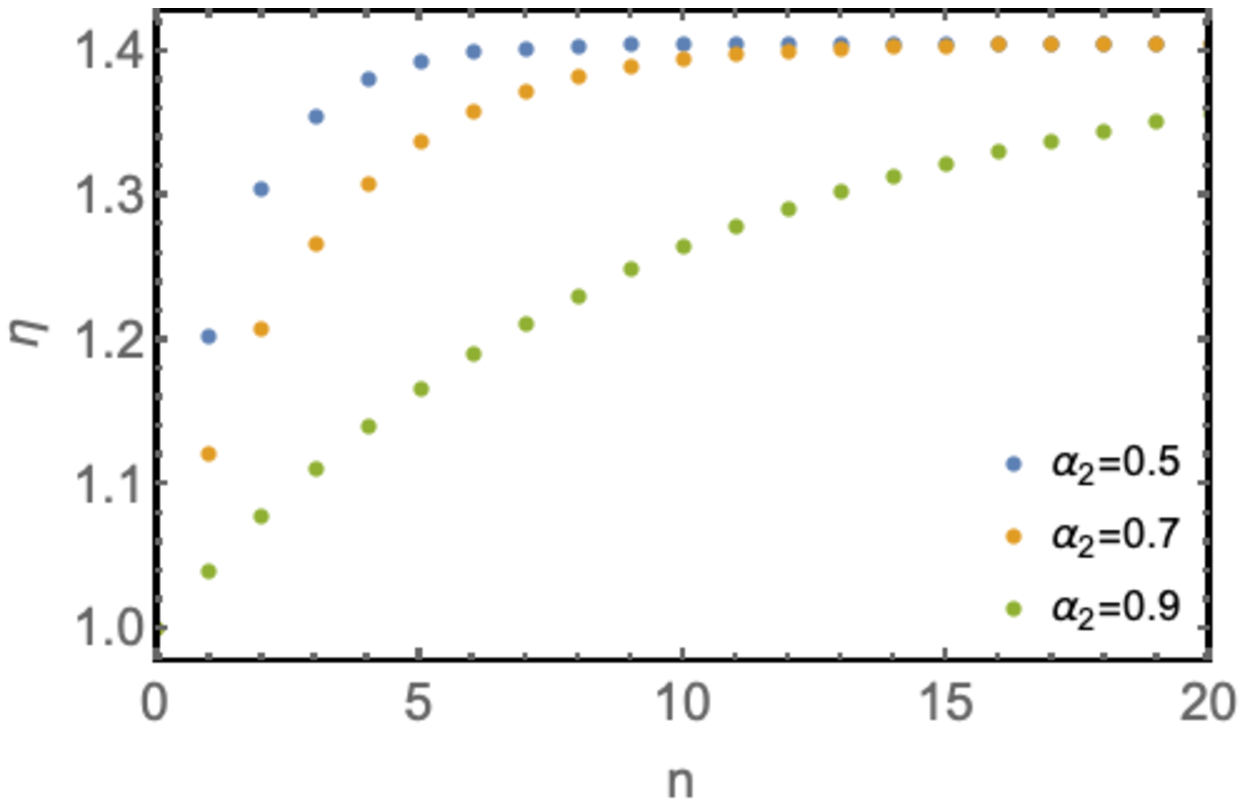}
          (a) 
        \end{center}
      \end{minipage}
      
      \begin{minipage}{0.48\hsize}
        \begin{center}
          \includegraphics[scale=0.63]{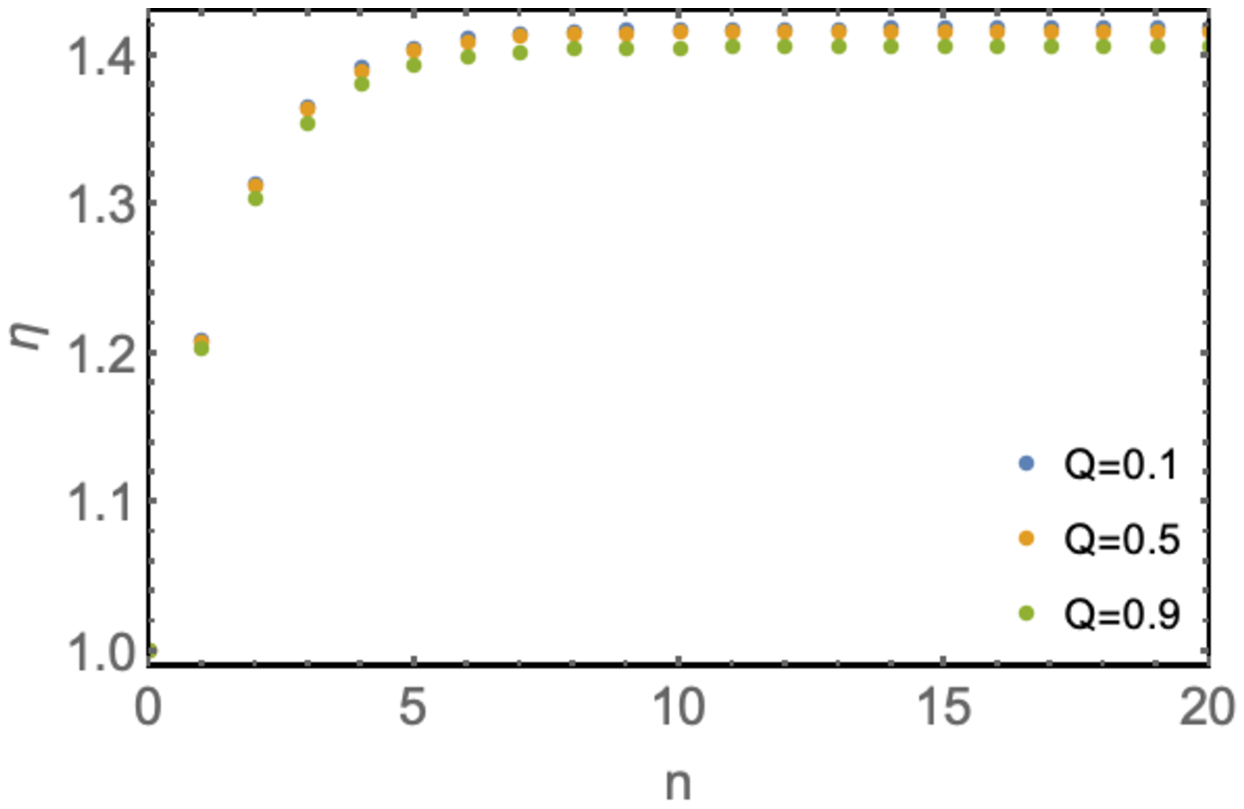}
         (b) 
        \end{center}
      \end{minipage}
\\      
      \begin{minipage}{0.48\hsize}
        \begin{center}
          \includegraphics[scale=0.63]{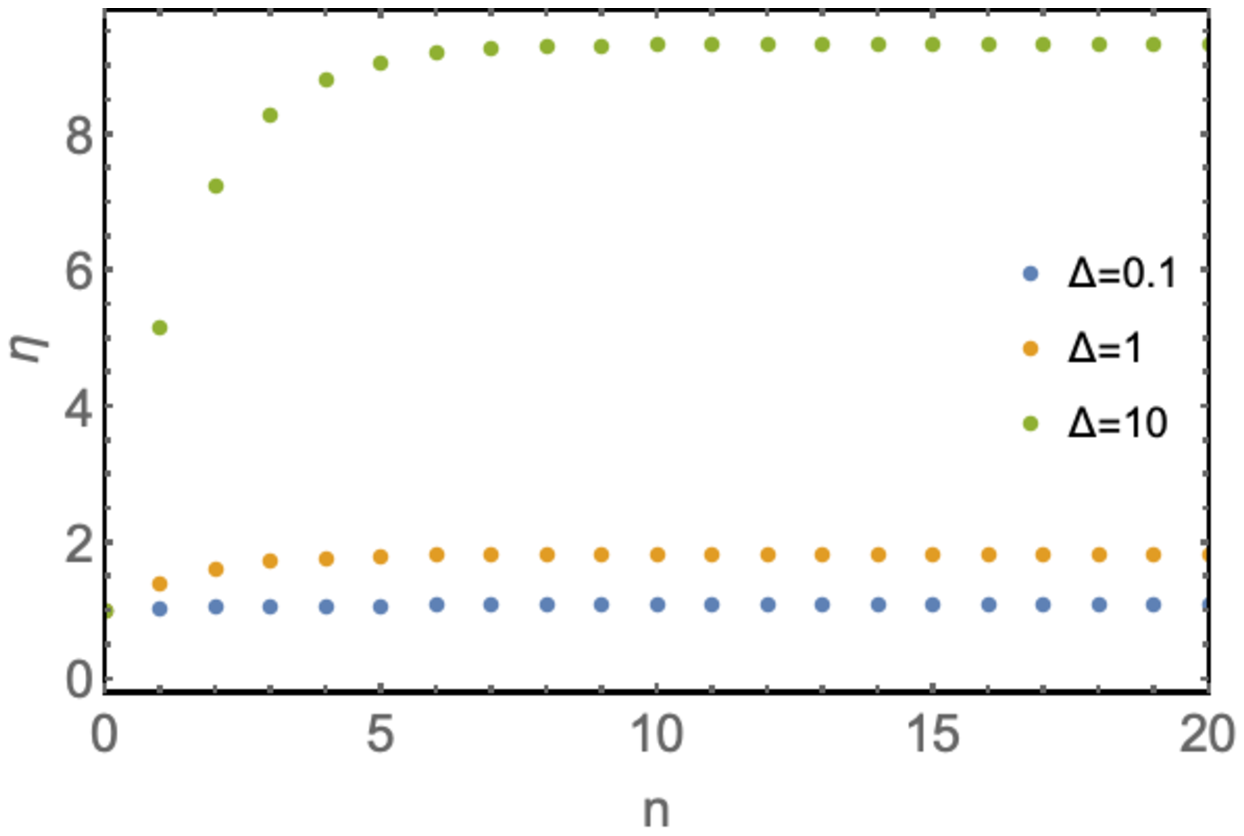}
         (c) 
        \end{center}
      \end{minipage}

      \begin{minipage}{0.48\hsize}
        \begin{center}
          \includegraphics[scale=0.63]{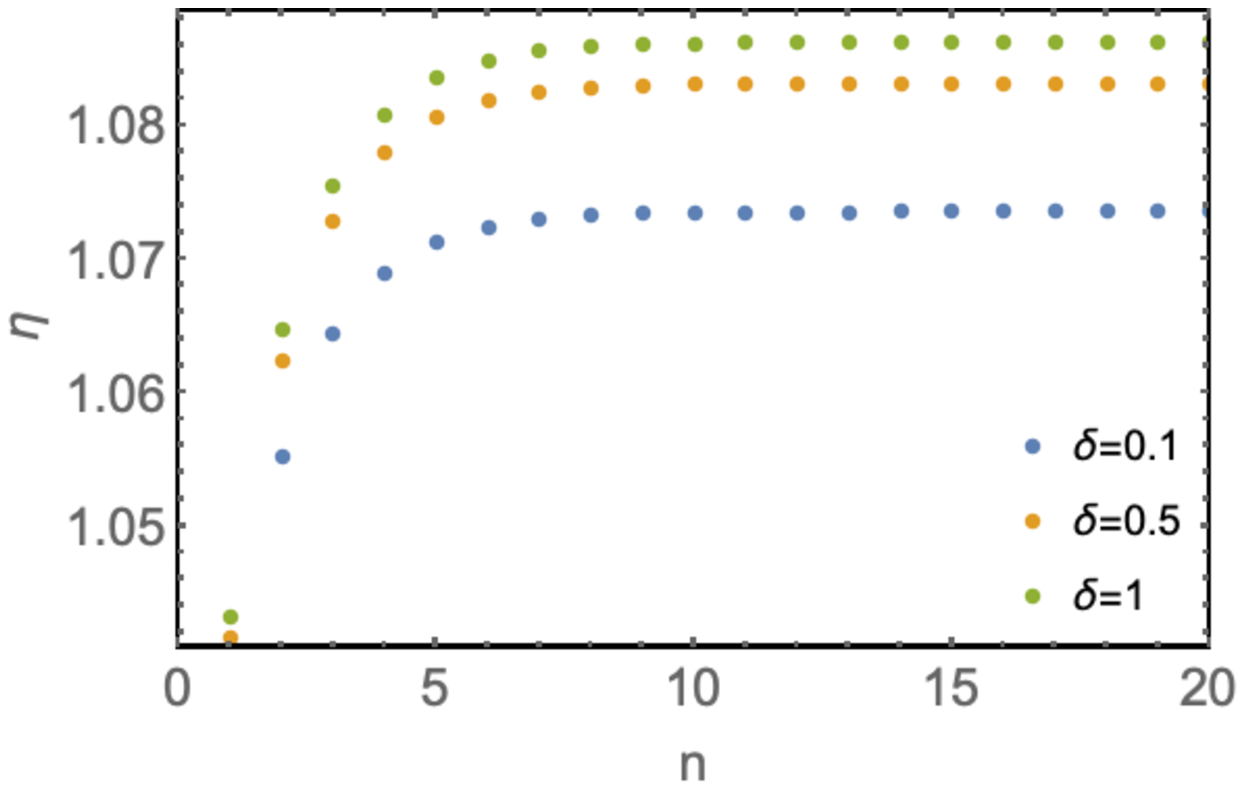}
         (d) 
        \end{center}
      \end{minipage} 
\\
    \end{tabular}
    \caption{(a) The energy amplification factor $\eta$ for first 20 decays under the parameters of $\Delta=0.5, \delta=0.5, E_0=1.1, Q=0.9$ and $\alpha_2=0.5, 0.7$ and $0.9$.
     (b) Same as (a) but for $\Delta=0.5, \delta=0.5, E_0=1.1, \alpha_2=0.5$ and $Q=0.1, 0.5$ and $0.9$. 
     (c) Same as (a) but for $ \delta=0.5, E_0=1.1,Q=0.5$ and $\Delta=0.1, 1$ and $10$.
     (d) Same as (a) but for $\Delta=0.5, E_0=1.1,Q=0.5$ and $\delta=0.1, 0.5$ and $1$.}    
\label{fig-eta}
  \end{center}
\end{figure}

Clearly, multiple processes can occur by repeating the above operation.
Summarizing, the amount of energy and charge for the particle, labeled as Particle $2n$, after $n$-th decays are calculated to be
\begin{align}
E_{2n}&=\left\{ E_0+(1-\alpha_2^n)\Delta \sqrt{f(x_s)} \right\}/\alpha_2^n, \label{E2n} \\
q_{2n}&=\left\{ q_0+(1-\alpha_2^n)(1+\Delta)\frac{x_s \sqrt{f(x_s)}}{Q} \right\}/\alpha_2^n. \label{q2n}
\end{align}
Therefore, we finally obtain $\eta$ after the $n$-th split,
\begin{align}
\eta=\frac{\tilde E_{2n}}{\tilde E_0}=\frac{\alpha_2^n E_{2n}}{E_0}
=1+(1-\alpha_2^n)\frac{\Delta \sqrt{f(x_s)}}{E_0}.
\label{eta-n}
\end{align}
As can be seen from \eq{eta-n}, $\eta$ increases at each decay and approaches a constant as the number of decay increases, i.e.,
\begin{align}
\lim_{n \to \infty}\eta=1+\frac{\Delta \sqrt{f(x_s)}}{E_0}. \label{saturation}
\end{align}
Thus, {\it the energy amplification considered here is not exponential, implying a stability of the confinement system at least in the test particle approximation.}
We plot $\eta$ for several parameters in \fig{fig-eta}.
From these figures it is obvious that the background charge $Q$ does not significantly affect $\eta$ . On the other hand, $\eta$ becomes unbounded if $\Delta$  takes a large value.

\begin{figure}[htbp]
  \begin{center}
    \begin{tabular}{c}

      \begin{minipage}{0.5\hsize}
        \begin{center}
          \includegraphics[scale=0.65]{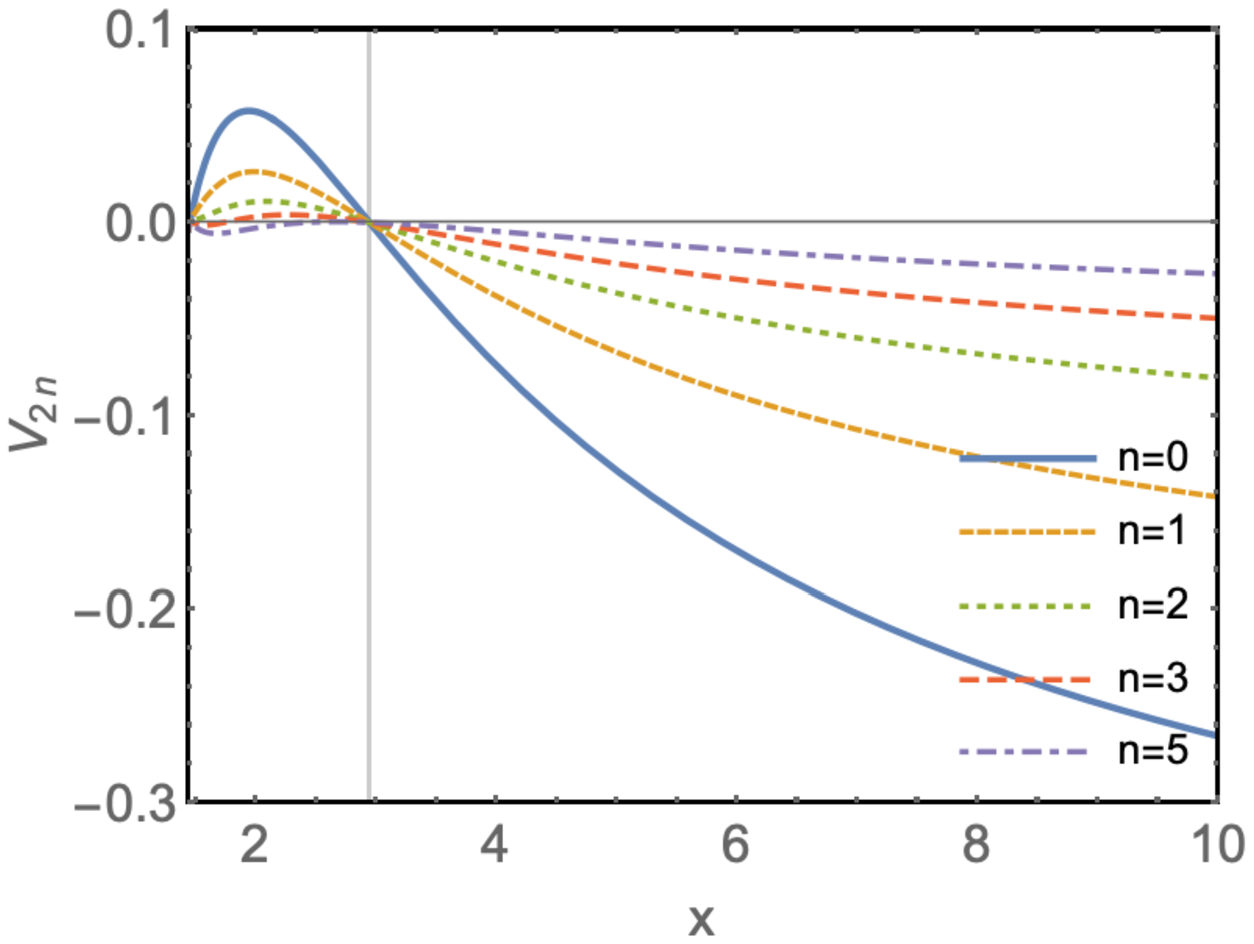}
          (a) 
        \end{center}
      \end{minipage}
      
      \begin{minipage}{0.5\hsize}
        \begin{center}
          \includegraphics[scale=0.7]{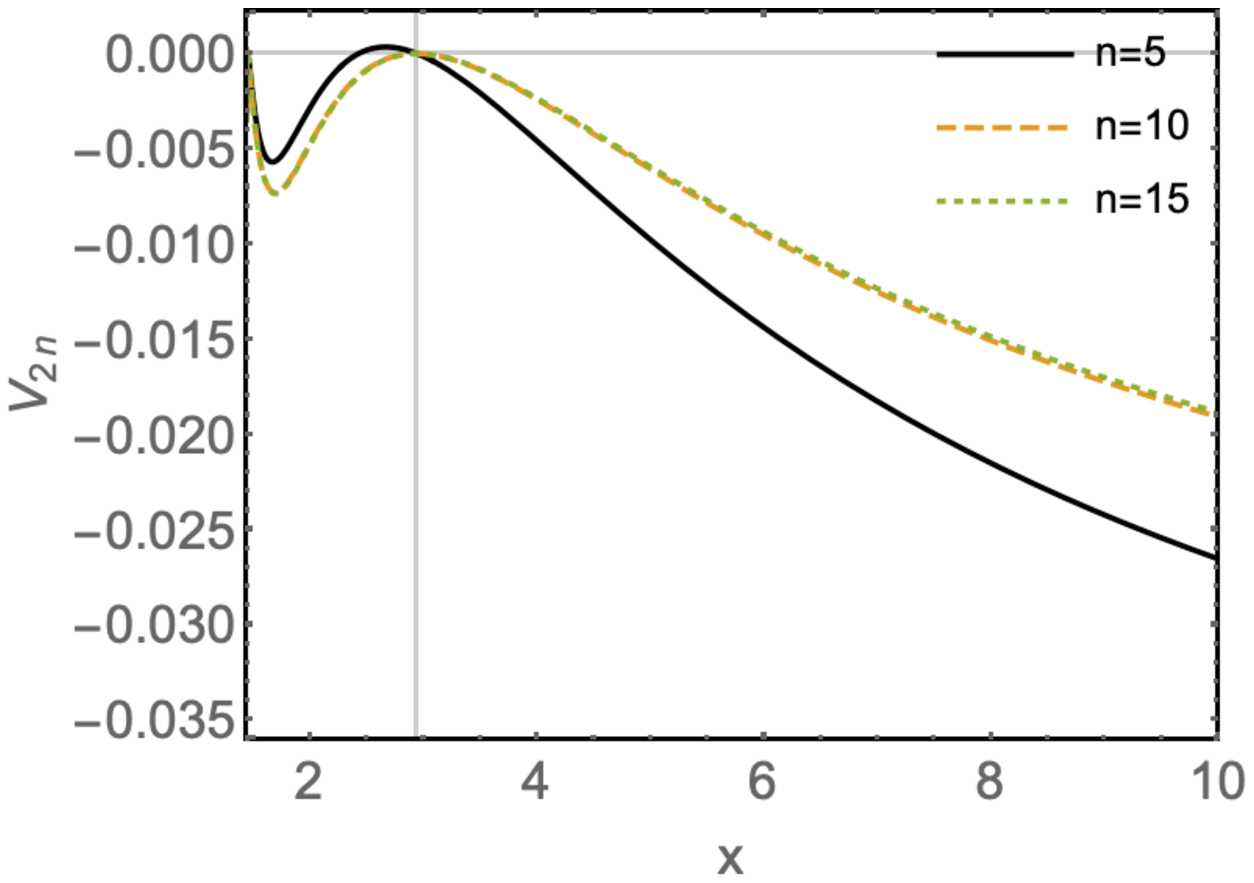}
         (b) 
        \end{center}
      \end{minipage}

    \end{tabular}
    \caption{Potentials of outgoing positive energy particles after the $n$-th decay. $n=0$ denotes the potential for Particle $0$. Particles splits at the same radius $x_s$, denoted as a vertical solid line. (a) Potentials under the parameter set of $Q = 0.9, E_0 = 1.2, \Delta = 0.1, \delta= 0.1, \alpha_2 = 0.5$ for $n=0, 1, 2, 3, 5$ (a) and for $n=5, 10, 15$ (b).
 }    
\label{fig-V2n}
  \end{center}
\end{figure}

The potentials of particles with energy Eq. (\ref{E2n}) and charge Eq. (\ref{q2n}) after the $n$-th decay, labeled as $V_{2n}$, are depicted in \fig{fig-V2n}.
It is clear from \fig{fig-V2n} that the energy and the velocity of the particle decreases after each decay. 
Since the mass of the fragment becomes smaller than the original particle ($0 <\alpha_i <1$), the motion of the fragment approaches closer to that of a photon with each decay.
The value of the potential $V_{2n}$ after {\it a large number of decays} is given by 
\begin{align}
\lim_{n \to \infty} V_{2n}=-\left(E_0-E_1-\frac{(q_0-q_1)Q}{x}\right),
\end{align}
which is in fact the same as the potential of a positive energy particle after {\it a single decay} in case the particle is a photon ($m \to 0$):
\begin{align}
\lim_{n \to \infty} V_{2n}=\lim_{m \to 0} V_2.
\end{align}
Consequently, we have
\begin{align}
\lim_{n \to \infty} \eta=\lim_{\alpha_1 \to 1} \eta |_{n=1}.
\end{align}
This relation means that, in terms of amount of the amplification,  ``amplifying energy gradually by timelike particles'' is equivalent to ``amplifying energy all at once by a null particle''.

\section{Summary and conclusion}
\label{sec:summary}
We have studied the charged Penrose process with radially moving particles with generic parameters.
We summarize our investigation below:
\begin{itemize}
\item If the split of particles is to occur under the assumption, $\dot r_{1,2}=0$, the mass of particles is conserved.

\item The negative energy of a fragment that falls into the BH can be arbitrarily close to zero, but the charge generally cannot, unless the background spacetime is extremal or the parent particle has a very large energy from the beginning.

\item The energy amplification for the charged Penrose process has no bound unlike the  process in the Kerr spacetime. An unbounded amount of energy is obtained by dropping infinitely large amount of negative charge.

\item The closer the mass of the escaping particle is to zero, the larger the energy amplification factor $\eta$. The maximum $\eta$ is achieved when the escaping fragment is a photon.

\item The potential of the fragment with  energy larger than that of the parent particle is qualitatively the same as the potential of the parent particle, which allows further decay to occur, once a reflective boundary is placed outside the horizon.

\item Our confinement system makes the particle's energy amplified each time a decay occurs. Thus, the energy is a monotonically increasing function of time. However, the energy asymptotes to a certain finite value given by \eq{saturation}. In other words, the energy amplification considered here is not exponential, implying a stability of the system. 

\item In terms of the value of $\eta$, the amplification rate is the same when the energy is gradually increased by multiple splits ( \eq{saturation}) and when the conversion into a photon is achieved  by a single split (\eq{eta} with $\alpha_1 \to 1$). In other words, amplifying energy gradually by timelike particles is equivalent to amplifying energy all at once by a null particle.

\end{itemize}

The energy amplifying phenomenon considered here looks similar to the charged BH bombs by the test scalar field \cite{Degollado:2013bha}, but they are different.
Since confined the test scalar field shows an exponential growth of its energy implying the linear instability,
it is fundamentally different from our result, and thus we cannot directly compare the test scalar field with our test particle in this respect.
Furthermore, our result seems also to be similar to a nonlinear instability of the charged BH bomb, but it is again different. The non-linear time evolution of the charged anti-de Sitter scalar field shows an exponential energy growth when the linear approximation is effective, but when the field evolution reaches a non-linear regime, the amplitude of the field saturates \cite{Bosch:2016vcp}. This saturation seems to be similar to ours, but since our system employs test particles where the back-reaction is not considered, our system is essentially different.
An exponential growth of energy may be possible in test particles by some other mechanism, but at least in the present model, such a growth does not occur.

In this study we considered a spherically symmetric spacetime for simplicity, but clearly, this analysis can also be performed in the Kerr background. In the Kerr case as well, the energy amplification may be caused by placing a reflective wall. When the initial velocity of fragments is zero, the energy amplification factor in the Kerr case is expected to gradually approach a constant value, as in the present study.

In this study, an artificial reflective boundary was placed, but it is naturally expected that energy amplification considered here will occur if there is a mechanism that causes confinement. Asymptotic anti-de Sitter spacetimes naturally provide such a confinement mechanism. The anti-de Sitter boundary provides a natural reflective wall in spacetime.
As yet another confinement mechanism, a system incorporating the self-gravity of particles can be considered. 
The self-gravity of matter (without pressure) also naturally provides a reflective boundary.
Israel's treatment \cite{Israel:1966rt} of the shell not only makes this possible, but also considers the back-reaction of the charge absorbed by the BH.  The shell model allows for a more complete analysis with these back-reactions.

Finally, we have assumed that two fragments are at rest at first for simplicity. But, this could be a strong assumption. Fragments can have in general non-vanishing velocities. In this general case, $\dot r_2 (x_s) \neq 0$, which means that the decay radius of the positive energy particle at the next decay is smaller than that of the first parent particle. As a result, the decay radius should gradually approach the horizon radius due to the continuous splitting of the particles in the confinement system.
It is interesting to evaluate the energy amplification in the general case.
This case is now under investigation.

\acknowledgments
T.K. is grateful to Masashi Kimura, Tomohiro Harada, Naoki Tsukamoto and Takuya Katagiri for fruitful discussions.
The authors are also grateful to anonymous referee for valuable comments. 
T. K., S. L., P. W. and H. Y. were supported in part by the NSFC under Grants No. 12005059, No. 12105098, No. 11947216, No. 11690034, No. 11805063, No. 11775077, and No. 12075084, and the China Postdoctoral Science Foundation (No. 2019M662785)


\end{document}